\documentclass{mn2e}
\usepackage[pdftex]{graphicx}

\def\ltsima{$\; \buildrel < \over \sim \;$}
\def\lsim{\lower.5ex\hbox{\ltsima}}
\def\gtsima{$\; \buildrel > \over \sim \;$}
\def\gsim{\lower.5ex\hbox{\gtsima}}

\newcommand{\be}{\begin{equation}}
\newcommand{\en}{\end{equation}}

\def\msole {~M_{\odot}}

\begin{document}
   \title[Variable Ly$\alpha$ in GRB 090426]{Variable Ly$\alpha$ sheds light on the environment
surrounding GRB 090426}

\author[C.C. Th\"one et al.]{C. C. Th\"one$^{1,2,}$\thanks{E-mail: cthoene@iaa.es}, 
S. Campana$^{1}$, D. Lazzati$^3$, A. de Ugarte  Postigo$^{1,4}$, J. P. U. \newauthor  Fynbo$^4$, 
L. Christensen$^5$, A. J. Levan$^6$, M. A. Aloy$^7$, J. Hjorth$^4$, P. Jakobsson$^8$,\newauthor
E. M. Levesque$^{9,10}$, D. Malesani$^4$, B. Milvang-Jensen$^4$, P. W. A. Roming$^{11,12}$,\newauthor
N. R. Tanvir$^{13}$, K. Wiersema$^{13}$, M. Gladders$^{14}$, E. Wuyts$^{14}$, H. Dahle$^{15}$\\
$^1$ INAF-Osservatorio Astronomico di Brera, Via Bianchi 46, I--23807, Merate (Lc), Italy\\
$^2$ Instituto de Astrof\'isica de Andaluc\'ia (IAA-CSIC), Glorieta de la Astronom\'ia s/n, 18008 Granada, Spain \\
$^3$ NC State University, Department of Physics, Riddick Hall, Raleigh, NC, USA\\
$^4$ Dark Cosmology Centre, Niels Bohr Institute, University of Copenhagen, Juliane Maries Vej 30, 2100 Copenhagen, Denmark\\
$^5$ Excellence Universe Cluster, Technische Universit\"at M\"unchen, Boltzmannstr. 2, 85748 Garching, Germany\\
$^6$ Department of Physics, Univ. of Warwick, Coventry, CV4 7AL, UK\\
$^7$ Fundaci\'o General and Departamento de Astronom\'ia y Astrof\'isica, Universidad de Valencia, 46100 Burjassot, Spain\\
$^8$ Centre for Astrophysics and Cosmology, Science Institute, University of Iceland, Dunhagi 5, 107 Reykjav\'ik, Iceland\\
$^9$ Institute for Astronomy, University of Hawaii, 2680 Woodlawn Dr., Honolulu, HI, 96822, USA\\
$^{10}$ Einstein Fellow, Center for Astrophysics and Space Astronomy, University of Colorado at Boulder, UCB 389, Boulder, CO 80309, USA\\
$^{11}$ Department of Astronomy and Astrophysics, PSU, University Park, PA 16802, USA\\
$^{12}$ Current Address: Space Science Department, Southwest Research Institute, San Antonio, TX 78238, USA\\
$^{13}$ Department of Physics and Astronomy, University of Leicester, University Road, Leicester, LE1 7RH, UK\\
$^{14}$ Department of Astronomy and Astrophysics, University of Chicago, 5640 S. Ellis Ave, Chicago, IL 60637, USA\\
$^{15}$ Institute of Theoretical Astrophysics, University of Oslo, P.O. box 1029 Blindern, N -- 0315 Oslo, Norway
}

\date{Accepted. Received.}

\pagerange{\pageref{firstpage}--\pageref{lastpage}} \pubyear{2010}

\maketitle

\label{firstpage}

\begin{abstract}
Long duration gamma-ray bursts are commonly associated with the deaths of
massive stars. Spectroscopic studies using the afterglow as a light source provide a unique
opportunity to unveil the medium surrounding it, probing the densest region of their
galaxies. This material is usually in a low ionisation state and at large distances from
the burst site, hence representing the normal interstellar medium in the galaxy. Here
we present the case of GRB 090426 at $z=2.609$, whose optical spectrum indicates an almost fully ionised
medium together with a low column density of neutral hydrogen. For the first time, we also observe variations in the Ly$\alpha$ absorption line.
Photoionisation modeling shows that we are probing material from the vicinity of the burst ($\sim 80$ pc).
The host galaxy is a complex of two luminous interacting galaxies, which might suggest 
that this burst could have occurred in an isolated star-forming region outside its host galaxy created in the interaction of the two galaxies.
\end{abstract}


\section{Introduction}

Optical afterglow studies of gamma-ray bursts (GRBs) provide a powerful tool for unveiling the interstellar medium (ISM) properties  of their host galaxies. Long GRBs are connected to the death of a massive star, and their host galaxies 
are subsequently expected to be sites of heavy star-formation extending back to the early Universe.
Metallicities determined from absorption lines of the ISM show higher values than those of, for example, QSO absorbers 
(Fynbo et al. 2006a; Savaglio et al. 2006; Savaglio 2009), though this might be only a sightline effect 
(Fynbo et al. 2008). Kinematics of absorption lines give some indications for galactic outflows 
(e.g. Th\"one et al. 2007), which are an expected phenomenon in star-forming galaxies.

Usually, absorption lines in GRB afterglow spectra are constant in time. In a few cases metal 
absorption lines together with their corresponding fine structure transitions, caused 
by UV pumping from the intense afterglow radiation, vary in intensity following the change in the 
ionizing flux of the burst. Modeling of the line variability allows us to better constrain the 
distance from the GRB to the absorbing material, which typically ranges from $>150$\,pc to a few kpc (e.g. Vreeswijk et al. 2007,  D'Elia et al. 2009). The presence of neutral material 
in the spectra implies distances on the order of a few hundred pc to several kpc (e.g. Prochaska et al. 2007, Th\"one et al. 2007). For strong NV absorption lines, it has been suggested that those lines must arise from material close to the GRB site, at $<$100\,pc (Prochaska et al. 2007). GRB 080310 showed variation in all resonant Fe II transitions which can only be explained with photoionization by the burst at a distance of $\sim$ 100\,pc (A. De Cia et al. in prep.). Excluding those few special cases, optical GRB afterglow spectroscopy seems to predominantly probe 
the ISM of the host galaxy and not the material around the GRB progenitor.

Studying the circumburst environment, however, would allow us to verify the predictions of current 
progenitor models. Prompt emission X--ray data showed some evidence for variations in the 
absorbing column density (Amati et al. 2000). In the optical, absorption features from circumburst 
material are difficult to detect due to the flash ionization of the surrounding material by the GRB 
out to several tens of parsecs (Robinson et al. 2010), even though the GRB is predicted to reside 
in a dense environment. In addition, any remaining absorbing material would be hidden under the 
stronger absorption lines from the line-of-sight ISM in the host galaxy.

There are several potential signatures which could demonstrate that we are observing 
material close to the GRB instead of the general ISM of the galaxy: time variability of the absorption line 
strength (Perna \& Loeb 1998), a high ionization state of the medium (Prochaska et al. 2008), 
or absorption lines at large velocities compared to the redshift of the GRB. The last has been 
observed in a few cases (Fox et al. 2008, M\o ller et al. 2002), but it cannot be excluded 
that those systems are intervening absorbers and hence not connected to the GRB. It was also suggested that a large difference between the optical and X-ray column densities might be a hint of observing highly ionized material, eventually close to the GRB (Watson et al. 2007, Campana et al. 2010). An ideal laboratory would be a GRB progenitor that exploded in a relatively isolated environment with little ISM along the line of sight. So far, no conclusive evidence for such conditions has been found, probably 
because such physical conditions are rare and only a few afterglows have time resolved afterglow 
spectroscopy. In this paper we present the first example for such a scenario, GRB 090426. 

GRB 090426 was discovered by the BAT (Burst Alert Telescope) $\gamma$-ray telescope onboard the
{\it Swift} satellite (Gehrels et al. 2004) on April 26, 2009, 12:48 UT (Cummings et al. 2009) and had a 
duration of $T_{90} = 1.2 \pm 0.3$\,s (Sato et al. 2009). Soon after its discovery 
an optical counterpart was found by {\it Swift} (Cummings et al. 2009) and the redshift was determined 
from an optical spectrum to be $z = 2.609$ (Levesque et al. 2010).
Both its restframe duration (0.3\,s) as well as its observed duration put GRB 090426 into the 
short burst category according to the commonly used classification based on the duration of the 
prompt emission. 

In \S 2 we describe the optical spectra and afterglow observations. In \S 3 we discuss the characteristics of GRB 090426 itself 
and some issues regarding its classification. 
\S 4 describes the properties of the ISM in the line of sight as derived from optical and X--ray 
observations. 
In \S 5 we deal in detail with the Ly$\alpha$ line variation and model its behaviour with a 
photoionisation code. \S 6 describes the properties of the host galaxy of GRB 090426. Discussion and conclusions
are reported in \S 7.

\noindent
\section{Observations}
We obtained both photometric and spectroscopic observations of the afterglow and the host galaxy. The imaging data are listed in Tab. \ref{logafterglow}, and the spectroscopic observations are detailed in Tab. \ref{log}. We observed the optical afterglow with MOSCA and AlFOSC at the NOT as well as FORS2 at the VLT from 2009 Apr. 26 to June 19 in several bands. In addition, we include acquisition images from the spectroscopic observations. Photometry was performed using SDSS stars (Abazajian et al. 2009) as reference and the transformations described by Jester et al. (2005) to convert the SDSS to V and R$_\mathrm{C}$ magnitudes. In order to account for the contribution of the host galaxy system to the light curve independent of the seeing, we used aperture photometry centered on the afterglow with radius of 2''.

We obtained a spectrum of the afterglow with FORS2 at the VLT starting on 2009 Apr 27.048 UT 
(12.3 hr after the burst) using grisms 600V and 600RI together with a 1\farcs0 slit and an exposure time of 30\,min per grism. 
The spectrum covers a wavelength range from 3500 to 8600\,\AA{} with a resolution of $\sim$ 11 \AA{}. 
Furthermore, we reanalyzed the spectrum from Levesque et al. (2010) taken with LRIS at Keck 1.1 hr after 
the burst. This spectrum has a similar resolution and wavelength coverage compared to the FORS2 spectrum. {\bf The spectra were reduced using standard tasks in IRAF, with the different exposures and grisms from each instrument combined, flux calibrated, and then normalized. Flux calibration prior to normalizing the spectrum makes the normalization more reliable, especially bluewards of Ly $\alpha$ and in the Ly $\alpha$ forest, since the afterglow continuum follows a smooth power law.}

After the fading of the afterglow an object was found $1\farcs1$ SW of the GRB 
position and initially suggested as the candidate host galaxy (G1). We took spectra of G1 with 
FORS2 at the VLT on June 18  2009 using grism 300V (3300 -- 8700 \AA{}) and exposure times of 
2\,$\times 1800$\,s. The acquisition image for the spectroscopic observations of G1 revealed another 
object underlying the GRB position (therefore suggested as the real host galaxy, H), for which 
we obtained another spectrum on June 19 2009 with the same settings and exposure times. We also 
obtained 2 $\times$ 30min spectra with X-shooter at the VLT (D'Odorico et al. 2006) on June 2 2009, covering a wavelength 
range from 9800 to 24800 \AA{}. The slit was placed across the GRB position (and the host) and G1.

\begin{table*}
\caption{Log of the imaging observations.}
\begin{tabular}{llllll}
\hline
Instrument   	& Start time (UT)    	& Time after GRB (d) & Exp. time (s) 	 & Band &Magnitude\\ \hline
MOSCA/NOT	& 2009 Apr 26.879	&  0.345  		&    3$\times$300      	  & R 	  &    21.27$\pm$0.16\\        
MOSCA/NOT	&2009 Apr 26.939   &  0.405 	&  1$\times$300            &R	&   21.33$\pm$0.11\\             
MOSCA/NOT	&2009 Apr 26.946  &  0.412 	&  2$\times$300               &i	& 21.18$\pm$0.02\\             
MOSCA/NOT	&2009 Apr 26.958  &  0.424	&  5$\times$200                &z	& 21.12$\pm$0.07\\               
MOSCA/NOT	&2009 Apr 27.015   & 0.481	& 2$\times$300                 &g	&21.86$\pm$0.03\\               
FORS2/VLT	&2009 Apr 27.048   &  0.514 	&  1$\times$120               &R &21.52$\pm$0.04 \\               
MOSCA/NOT	&2009 Apr 27.113  &  0.579 	&  2$\times$300               &g	& 22.12$\pm$0.05\\              
MOSCA/NOT	&2009 Apr 27.122  &  0.588   	&  2$\times$300             &V	& 22.05$\pm$0.06\\               
MOSCA/NOT	&2009 Apr 27.128   &  0.594  	&  1$\times$300             &R	& 21.77$\pm$0.12\\               
FORS2/VLT	&2009 Apr 27.144    &  0.610 	&  1$\times$120             &R	&21.92$\pm$0.04\\               
FORS2/VLT	&2009 Apr 27.150  &  0.616  	&  1$\times$120              &R	&21.95$\pm$0.04\\               
FORS2/VLT	&2009 Apr 27.152  & 0.618 	&  1$\times$60                  &R	&21.87$\pm$0.04 \\               
MOSCA/NOT	&2009 Apr 27.883  &  1.349   	&  3$\times$300             &R	& 22.85$\pm$0.19\\               
MOSCA/NOT	&2009 Apr 28.159  &  1.625 	&  4$\times$300               &R	& 23.00$\pm$0.13\\               
MOSCA/NOT	&2009 Apr 29.067   &  2.533  	&  6$\times$600              &R	&23.30$\pm$0.11\\               
AlFOSC/NOT	&2009 May 03.918  &  7.384	&   10$\times$600            &R	& 23.73$\pm$0.19\\               
FORS2/VLT	&2009 Jun 18.981  &  53.447   	&  2$\times$300            &V	&  23.76$\pm$0.17\\               
AlFOSC/NOT	&2009 Jun 19.915  &  54.381	&  7$\times$600               &R	&  23.71$\pm$0.17\\               
FORS2/VLT	&2009 Jun 19.989  & 54.455	&  300+450          &V	&  23.88$\pm$0.17\\               
FORS2/VLT	&2009 Jun 19.976  &  54.442	&  1$\times$300               &R	&  23.71$\pm$0.31\\               
\hline
\end{tabular}
\label{logafterglow}
\end{table*}

\begin{table*}
\caption{Log of the spectroscopic observations.}
\begin{tabular}{lllll}
\hline
Instrument   & Start time (UT)    & Exp. time (s) & Grism    & Resolution\\ 
\hline
Keck/LRIS    & 2009 Apr 26.506  & 2$\times$300     & 300/5000       & 650 \\
VLT/FORS2    & 2009 Apr 27.056& 2$\times$1800        & 600B & 740 \\
VLT/FORS2    & 2009 Apr 27.100& 2$\times$1800       & 600RI & 740 \\
VLT/X-shooter& 2009 May 31.975  & 3$\times$1200    & VIS+NIR   & $\sim$6000 \\
VLT/FORS2    & 2009 Jun 18.997 & 2$\times$1800        & 300V  & 440 \\
VLT/FORS2    & 2009 Jun 20.011 & 2$\times$1800       & 300V  & 440\\
\hline

\end{tabular}
\label{log}
\end{table*}

\section{GRB 090426: short or long?}

The original division of GRBs into two populations dates back to Kouveliotou et al. (1993) who found 
a bimodal distribution in the high-energy duration and hardness ratio (HR) of BATSE bursts. This was thought to imply two different progenitor populations, in particular since collapsars cannot produce bursts with a duration shorter than a few seconds. 
A third, intermediate, population was also proposed on statistical grounds (Horvath et al. 2010), 
but there is no conclusive evidence from properties of the bursts, afterglows, or host 
galaxies that this third class is intrinsically different from long bursts (de Ugarte Postigo et al. 2010). 
In recent years, the classification of GRBs based only on their high-energy properties has been 
widely discussed after the discovery of outliers such as long GRBs without detected supernovae 
(SNe) and a number of high-redshift bursts that are short in their rest-frame duration but likely coming from the collapse of a massive star. Levesque et al. (2010) statistically placed GRB 090426 in the short burst 
category; however, due to other properties they did not exclude a collapsar progenitor model.

\begin{figure}
%\begin{center}
\vspace{3cm}
\includegraphics[width=\columnwidth]{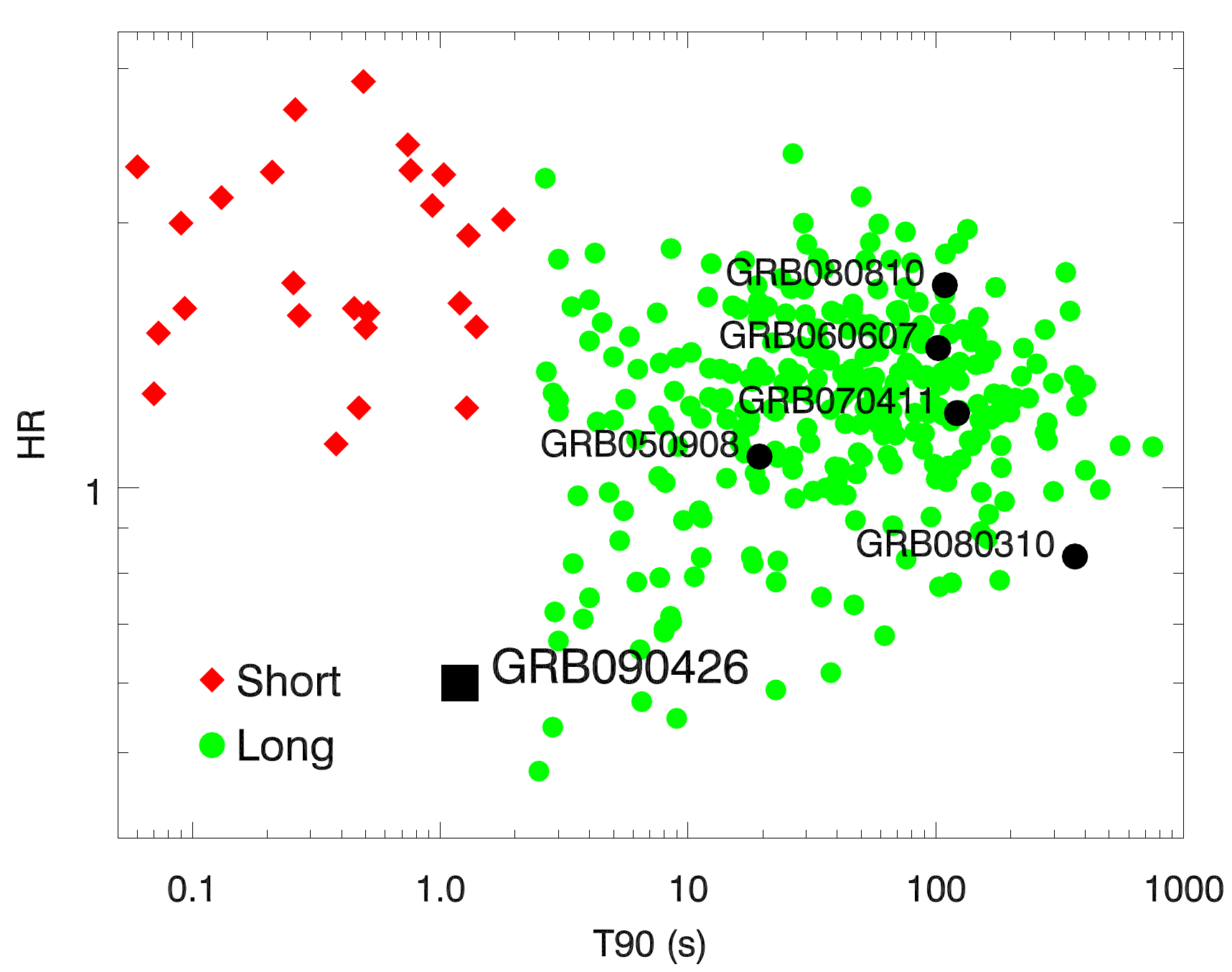}
\caption{ $T_{90}$ vs. hardness-ratio according to the BATSE channels (50-100 keV vs. 15-25 keV) applied to {\it Swift} 
bursts with known redshift (Horvath et al. 2010). Bursts marked in black are those 
with a column density of neutral hydrogen determined from optical spectra  with log N$_\mathrm{HI}/\mathrm{cm}^{-2} < 20$.}
\label{090426:HR}
%\end{center}
\end{figure}

In the following, we present some properties of the burst and afterglow that suggest it was 
likely produced by a collapsar event. 
In order to compare GRB 090426 with the BATSE bursts we transform the count ratio of the 
{\it Swift}-BAT data to those of the BATSE channels and obtain a ratio of (50-100 keV) / (15-25 keV) 
= 0.60, which classifies it as a ``soft'' burst. This places GRB 090426 in a very exceptional 
region of the HR vs. $T_{90}$ diagram, outside of the two main groups of long-soft and short-hard 
bursts with a spectrum that is even softer than most long-duration {\it Swift} bursts (see Fig. 
\ref{090426:HR}). The luminosity of its prompt emission with $E_{\rm iso} = (5\pm1)\times 10^{51}$ erg 
(1--10 000 keV, isotropic energy release, Antonelli et al. 2009) would be considered rather high for a merger event. {\bf However, it could be possible to reach such high luminosities through merger events. Recent simulations of neutron star-black hole mergers (Pannarale, Tonita \& Rezzolla 2010) show that the torus created in the merging process can approach or even exceed one solar mass, which would yield enough energy to power this burst. A high energy release could also be facilitated by neutrino-powered outflows or magnetic fields (e.g. Aloy, Janka \& M\"uller, 2005). }

%Such a high energy release could be facilitated by neutrino-powered outflow or magnetic fields \cite{??}. 
Another distinction between long and short burst is the determination of a spectral 
lag between the high energy emission in different bands, which is usually present for long bursts 
but absent for short bursts (Norris et al. 2006). For GRB 090426, the lag is consistent 
with zero but has a large uncertainty (Ukwatta et al. 2009).

\begin{figure}
\begin{center}
\includegraphics[width=\columnwidth]{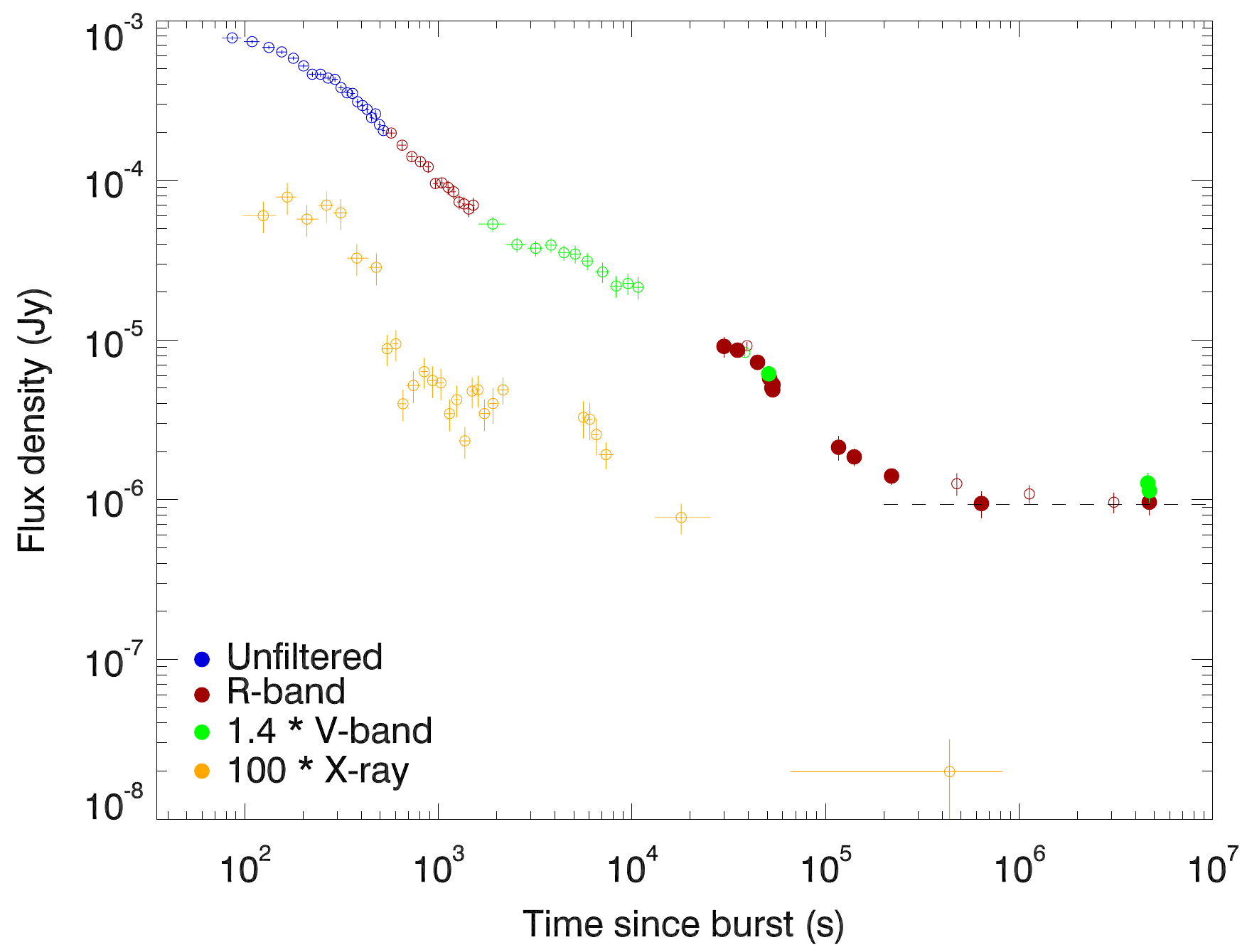}
\caption{Optical lightcurve collected from data in the literature (Antonelli et al. 2009; 
Xin et al. 2011; open symbols) and observations from the NOT and VLT (published in this article; filled symbols), all normalized to the $R$-band. The dashed line indicates the flux contribution from the underlying host galaxy. We plot the X-ray lightcurve for comparison with the flux density provided by the {\it Swift} archive (Evans et al. 2010).}
\label{090426:LC}
\end{center}
\end{figure}

The lightcurve of the optical afterglow was previously discussed in Antonelli et al. (2009) 
and Xin et al. (2011). We collect the lightcurve data available from the literature from 90 s to 35 d after the burst and add the observations listed in Tab. \ref{logafterglow}. We normalize all optical data to the $R$-band to construct a single lightcurve. The optical light cuve of 
GRB 090426 is similar to what is usually found for other long (or short) GRB afterglows. It reaches its maximum before 100 s, when the
observations published by Xin et al. (2011) begin, which likely marks the onset of the afterglow. Between 300 and 3,000\,s the light curve decays with a temporal slope of $1.12\pm0.02$.  At 3,000\,s we see a flattening in the light curve after which the decay seems to continue with approximately the same slope as before. This
bump could be due to an energy injection (Bj\"ornsson et al. 2004), as it is achromatic and happens simultaneously in optical
and in X-rays (see Fig. \ref{090426:LC}). Similar behaviour has been seen in a number of other bursts e.g. GRB 021004 (de Ugarte Postigo et al. 2004) and GRB 060526 (Th\"one et al. 2010). There are also indications of similar bumps at 10,000  
and 30,000\,s. After the first day, the lightcurve flattens when the light of the host galaxy at $R \sim 23.8$ mag 
begins to dominate. In Fig. \ref{090426:LC} we show the optical together with the X-ray lightcurve in flux units. The optical data were not corrected for the contribution from the host galaxy.

Kann et al. (2010) found a clear bimodality in the distribution of optical 
afterglow luminosities, with an average absolute magnitude of M$_\mathrm{B}$= 
--23.45 mag for long/type II GRBs and  M$_\mathrm{B}$= --18.2 mag for short/type I GRBs at 1 day after the burst (restframe).  {\bf This might be due to differences in the density of the surrounding material, which should be higher for long bursts that are typically found in dense star-forming regions. With an absolute magnitude of 
M$_\mathrm{B}$= --23.79 mag, the luminosity of the afterglow of GRB 090426 is around the average 
for long GRB afterglows, and about 5 magnitudes brighter than a typical short GRB. It has also been suggested that some compact object mergers have a very short coalescence time while they are still within their star forming region, and hence should have afterglow luminosities comparable to long bursts (Belczynski et al. 2006). The relatively high redshift of this burst compared to other short bursts would support this possibility.}

As previously suggested by other authors, we consider this burst to be peculiar. However, the properties detailed above place it 
closer to the population of GRBs originating from massive progenitors.

%A fit to the X--ray spectrum gives an absorbing column density of log $N_\mathrm{HI}$/cm$^{-2}=$  
%21.7$^{+0.3}_{-0.6}$ close to the average X--ray column density for long {\it Swift} GRBs (Campana et al. 2010).

\section{Properties of the ISM in the line-of-sight}

\subsection{X--ray spectra analysis}

We analysed seven observations with the XRT instrument onboard the {\it Swift} satellite, covering a time period from 
26 April to 5 May 2009. The source was only detected during the first two observations on 26 April 2009. 
The XRT data were reprocessed with standard procedures (xrtpipeline ver. 0.12.4 within the
Heasoft package ver. 6.8), resulting in 4.7 and 2.0 ks exposures in photon counting mode, respectively.
We decided to group the events to one photon per channel and to adopt Cash-statistics (Cash 1979)
within the fitting program XSPEC.
At the beginning of the first observation the source is slightly piled-up (two photons fall on the same
pixel during the same integration time and are read as a single photon with higher energy).
We take a conservative pile-up threshold of 0.7 ct s$^{-1}$ and split the first observation into
a piled-up interval (1.8 ks) and a non-piled-up interval (2.9 ks) which was summed to the second observation 
for a total time of 4.9 ks. Photons from the piled-up interval were extracted from an annular region with 
2 and 25 pixel inner and outer radii centered on source. Photons from the non-piled-up observations were 
extracted from a 20 pixel radius circular region. The background was extracted from the same annular region 
with an inner and outer radius of 60 and 80 pixels, respectively, both of which are free of field sources.

Ancillary response files were generated with the {\tt xrtmkarf} task, accounting for CCD defects,
vignetting, and point-spread function corrections using the appropriate exposure map.
We end up with two spectral files containing 383 and 328 counts in the 0.3--10 keV energy band.
The background is very low, comprising of less than $1\%$ and $7\%$ for the two observations, respectively.
We group the events and adopt Cash-statistics (Cash 1979) within the fitting program {\tt XSPEC}.
During the spectral fit we used the newly released broadened response file {\tt swxpc0to12s6\_20070901v011.rmf}.
The spectral model consists of a power law model and two absorption components, one from our Galaxy fixed to
$N_H=1.5\times 10^{20}$ cm$^{-2}$ (Kalberla et al. 2005) and one at the redshift of the host galaxy
($z=2.609$).

We find consistent spectral results between the two spectra, so we fit the column densities and the power 
law photon indices together and leave only the normalizations free to vary.
Assuming solar metallicity and solar composition, the best fit column density is $N_H=5.0^{+4.1}_{-3.8}
\times 10^{21}$ cm$^{-2}$ and the power law photon index is $\Gamma=2.1\pm0.1$ (errors are at $90\%$ 
confidence level with $\Delta\chi^2=2.71$)s. 
The column density is around the average found for other long {\it Swift} GRBs (Campana et al. 2010).

%\begin{figure}
%\begin{center}
%\includegraphics[width=\columnwidth]{Xraycontour.pdf}
%\caption{Contour plot of the allowed range for the X--ray absorption column density versus the 
%power law photon index. The contours are 1, 2 and 3 $\sigma$ confidence level.}
%\label{090426:Xray}
%\end{center}
%\end{figure}

\subsection{The optical spectrum}\label{spectrum} 

The spectrum of the optical afterglow turns out to be very different from most other GRB spectra. 
In our FORS spectrum we only detect very strong high ionization lines of N\,V $\lambda\lambda$ 1238, 1242, Si\,IV 
$\lambda\lambda$ 1393, 1402, and C\,IV $\lambda\lambda$ 1548, 1550. In addition, the Ly$\alpha$ 
absorption is relatively weak; however, this is not unprecedented in GRB afterglow spectra. In contrast to the usually strong low ionization 
lines detected in other GRB spectra, we only detect a weak C\,II $\lambda$ 1334 line while normally prominent 
lines of Si II $\lambda\lambda$ 1260, 1526, Fe\,II and Al\,II are undetected. This implies that the observed medium 
is in a high ionization state, which is very unusual. 

\begin{figure}
 \begin{center}
\includegraphics[width=\columnwidth,angle=-0]{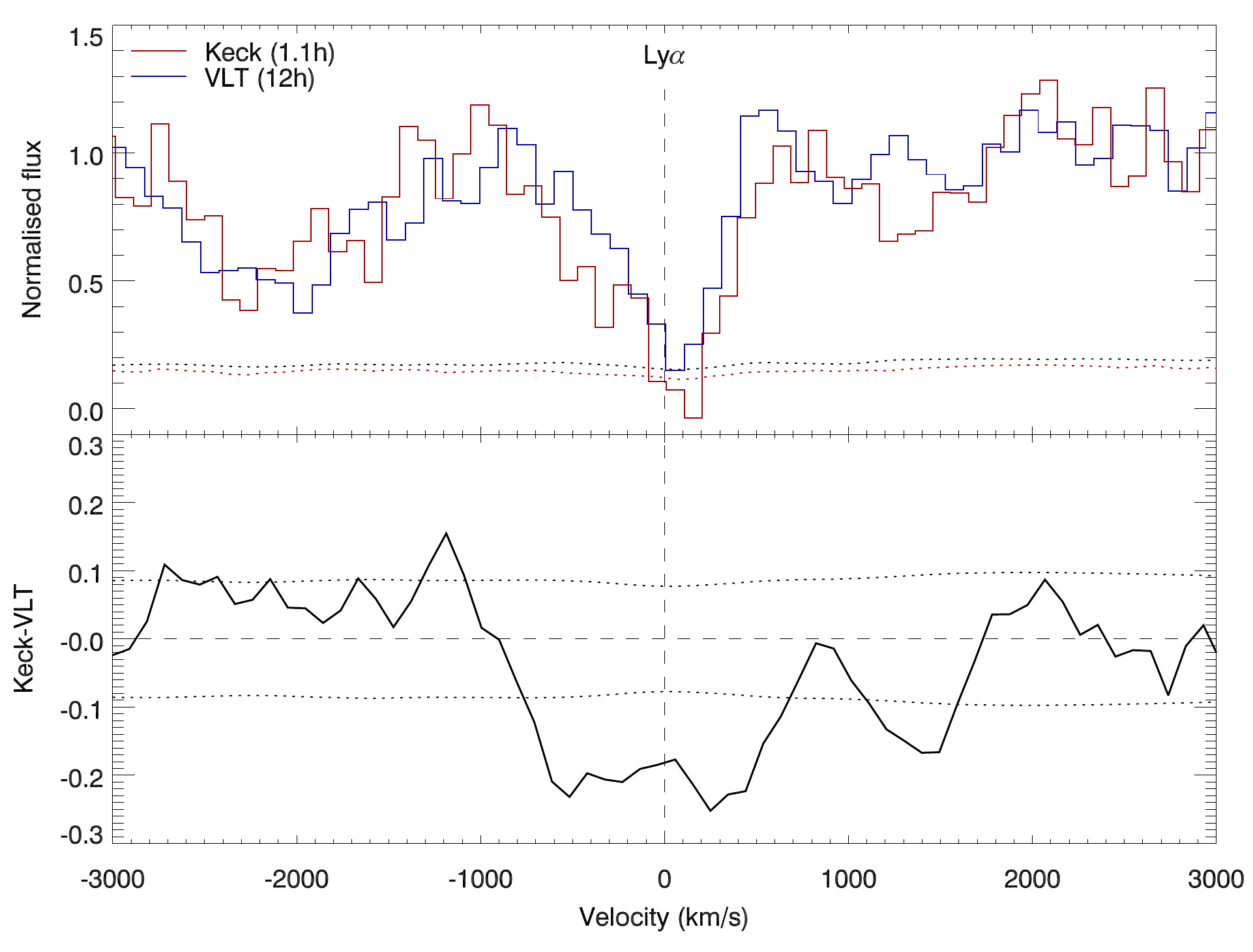}

\caption{Top panel: Ly$\alpha$ absorption line at two different epochs; the dotted lines are the 
corresponding error spectra. Bottom panel: Difference between the two epochs (solid line) and 1 $\sigma$ 
errors (dotted lines). The spectra have been smoothed with a running window before subtraction, corresponding to the instrumental width.}
   \label{090426:lyadiff}
 \end{center}
\end{figure}

We then compared our FORS spectrum from 12\,h after the burst with the the spectra by Levesque et al. (2010) 
from 1.1\,h after the burst. We detect the same high ionization absorption lines, but the Ly$\alpha$ 
line is significantly stronger than in our later spectra. We measure a 3.8 $\sigma$ decrease in the 
equivalent width (EW) between the two spectra at 1.1 and 12\,h after the GRB (see Fig. \ref{090426:lyadiff}). 
The other absorption lines do not show any significant variability. Variability of Lyman $\alpha$ has never 
been before observed in a GRB afterglow spectrum, nor in any other extragalactic object.

The column densities were obtained from the curve-of-growth (CoG)\footnote{The CoG describes the evolution of the column density with EW, with a linear relation for unsaturated lines and a dependence on the broadening of the line by turbulence or temperature in the following, flat part of the curve.}, adopting a $b$-parameter of 50 km\,s$^{-1}$. 
The errors on the Ly$\alpha$ column densities only reflect the measured errors on the EWs, not the 
uncertainty in the $b$-parameter, which would change the absolute values for the column densities but 
not the significance of the temporal variation. For the other lines, we assume very conservative 
errors on the column densities to account for the uncertainty on the $b$-parameter of around 
15 km\,s$^{-1}$. Si\,IV has two components resolved and we therefore determine the total column density 
by summing the contributions from the individual components (which are each closer to the linear part 
of the CoG than the combined line). For C\,IV, the column density was calculated assuming an EW ratio 
of 1:2 for the blended 1549/1550 doublet and taking the column density from the CoG for the weaker 
transition (1550\,\AA{}). The column density listed is therefore not for the doublet together but 
closer to the ``true" column density (had it been possible to deblend the two lines). However, 
since the line is clearly saturated, the true ratio deviates from the theoretical 1:2 ratio and the 
error in the column density must therefore be considered much larger.

\begin{figure*}
\begin{center}
\includegraphics[width=14cm]{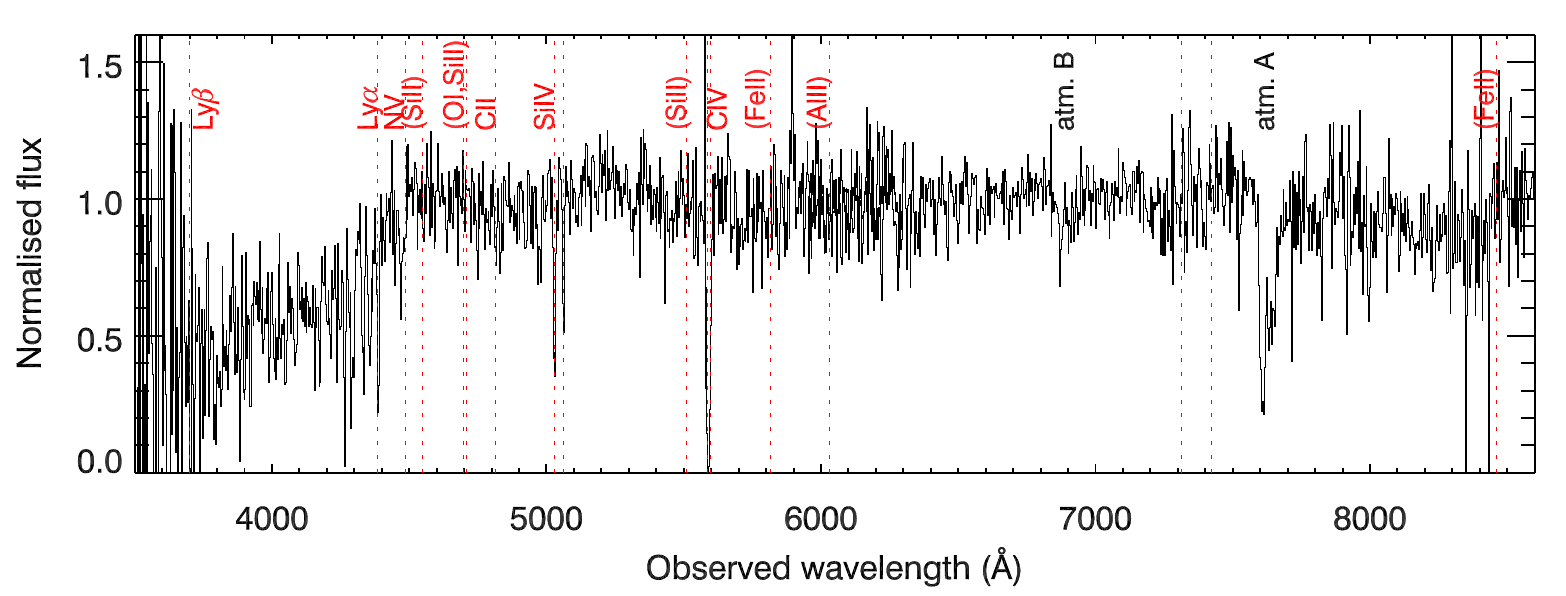}
\caption{Normalized spectrum from FORS2/VLT grism 600B and 600RI with absorption lines indicated. 
Undetected lines that are commonly detected in other GRB spectra are written in brackets. Atm. A and B mark the major atmospheric absorption bands.}
\label{090426:spec}
\end{center}
\end{figure*}

In Table 3 we list the measured equivalent widths (EWs) and derived column densities for the VLT/FORS 
spectrum and the comparison with the spectrum presented in Levesque et al. (2010). We also list the 
deviation of the individual lines between the two epochs. Both spectra have been analysed consistently 
by our group, and the EW values might therefore slightly deviate from those listed in Levesque et al. (2010). 
A plot of the FORS2 spectrum is shown in Fig. \ref{090426:spec}.

\begin{table*}
 \centering
 \begin{minipage}{140mm}
\caption{Absorption line properties and upper limits on some transitions that are usually strong in 
other GRB spectra but undetected here. EWs are given in restframe.}
%  \begin{tabular}{@{}llrrrrlrlr@{}}
\begin{tabular}{lllllll} \hline
Ion	     &$\lambda_\mathrm{rest}$&	$\lambda_\mathrm{obs}$	&EW$_\mathrm{VLT}$ (12\,h)	&log $N$	&	EW$_\mathrm{Levesque}$ (1.1h) 	&EW diff.\\ 
             & (\AA{})&	(\AA{})&(\AA{})&(cm$^{-2}$)&(\AA{})& \\ \hline\hline			
H\,I         &1215		        &	4385			&1.94$\pm$0.23		&18.7$^{+0.1}_{-0.2}$&	3.09$\pm$0.2	&--3.8$\sigma$\\
	     &				&				&					&				&(log N$_\mathrm{HI}=19.2\pm0.1$)&	\\
N\,V	     &1238			&	4471			&0.99$\pm$0.20		&15.8$\pm$0.8	&	0.70$\pm$0.10	&---\\
N\,V	     &1242			&	4485			&0.29$\pm$0.07		&15.8$\pm$0.8	&	0.30$\pm$0.10	&---\\
Si\,II/S\,II &1260			&	5457			&$<$ 0.35		& $<$ 13.6	&	$<$ 0.35	&---\\
Si\,II	     &1304			&	4797			&$<$ 0.25		& $<$ 13.6	&	---		& ---\\
C\,II	     &1334			&	4814			&0.86$\pm$0.19		& 15.8$\pm$1.0  & 	0.53$\pm$0.14	&+1.3$\sigma$\\
Si\,IV	     &1393			&	5030			&2.09$\pm$0.20		& 15.4$\pm$1.0  &	1.78$\pm$0.15	& ---\\
Si\,IV	     &1403			&	5062			&1.38$\pm$0.20		& 15.4$\pm$1.0  &	1.53$\pm$0.14	&--1.5$\sigma$\\
Si\,II	     &1526			&	5504			&$<$ 0.35		& $<$ 13.6	&	---		&---\\
C\,IV	     &1549			&	5591			&4.32$\pm$0.28		& 17.6$\pm$1.0  &	3.63$\pm$0.25	&+1.8$\sigma$\\
Fe\,II	     &1608			&	5804			&$<$ 0.34		& $<$ 14.4	&	---		&---\\
Al\,II	     &1670			&	6029			&$<$ 0.37		& $<$ 12.9	&	---		&---\\ \hline
\end{tabular}
\end{minipage}
\end{table*}

The variability of Ly$\alpha$ strongly suggests that the material lies 
close to the GRB progenitor. Such an effect has been predicted from photoionization modeling 
(Perna \& Loeb 1998), but has never before been observed.
The high column densities of the high ionization lines and the absence of their low ionization 
counterparts support the scenario that we are observing a large quantity of matter close 
to and ionised by the GRB. This is further supported by the large difference between the optical and 
X--ray column density, which can be explained by photoionization of the medium by the GRB; X--ray absorption is insensitive 
to this phenomenon since it is caused by excitation of the electrons in the inner shells of the atoms 
(Campana et al. 2010).

\subsubsection{Comparison to the ``average'' GRB spectrum}

We compare the lines found in the spectrum of GRB 090426 to a sample of 77 low-resolution afterglow 
spectra of long GRBs observed by {\it Swift} (Fynbo et al. 2009). The majority of those spectra in the 
right wavelength range to detect Ly$\alpha$ absorption (33) showed a strong damped Ly$\alpha$ (DLA) 
system with a median value of log N$_\mathrm{HI}$/cm$^{-2}$ $=$ 21.6. Only 6 other 
spectra had column densities below log N$_\mathrm{HI}$/cm$^{-2}$ = 20.0, the lower limit for the 
definition of a DLA: GRB 050908 (log N$_\mathrm{HI}$/cm$^{-2}$ = 17.8), GRB 060124 
(log N$_\mathrm{HI}$/cm$^{-2}$ = 18.5), GRB 060607A (log N$_\mathrm{HI}$/cm$^{-2}$ $\sim$ 17.2), 
GRB 070411 (log N$_\mathrm{HI}$/cm$^{-2}$ = 19.3), GRB 071020 (log N$_\mathrm{HI}$/cm$^{-2}$ $<$ 20.0), 
and GRB 080310 (log N$_\mathrm{HI}$/cm$^{-2}$ = 18.8). All of them securely belong to the 
long burst group with $T_{90}$s between 25 and 100\,s. Five of these bursts are also highlighted 
in Fig. \ref{090426:HR}. 

Christensen et al. (2010) produced a composite using 60 of the 77 spectra from Fynbo et al. (2009) mentioned above, and determined EWs and column densities for all detected lines in the composite. The following lines, which are usually strong in the environments of long-duration GRBs, are not detected in our spectrum: SiII  
$\lambda$ 1260,1304 and 1526 \AA{}, C II  $\lambda$1534,  and Al III $\lambda$ 1670,1855 and 1863.  
The EWs of C IV and Si IV in the spectrum of GRB 090426 are almost twice as high as in the composite 
GRB spectrum. Considering that the HI column density of GRB 090426 is very low compared to the 
approximate value for the composite (log $N_\mathrm{HI}$/cm$^{-2}$= 21.7), the relative abundance of 
these species are rather extreme for this burst. 

\begin{figure}
\begin{center}
\includegraphics[width=\columnwidth]{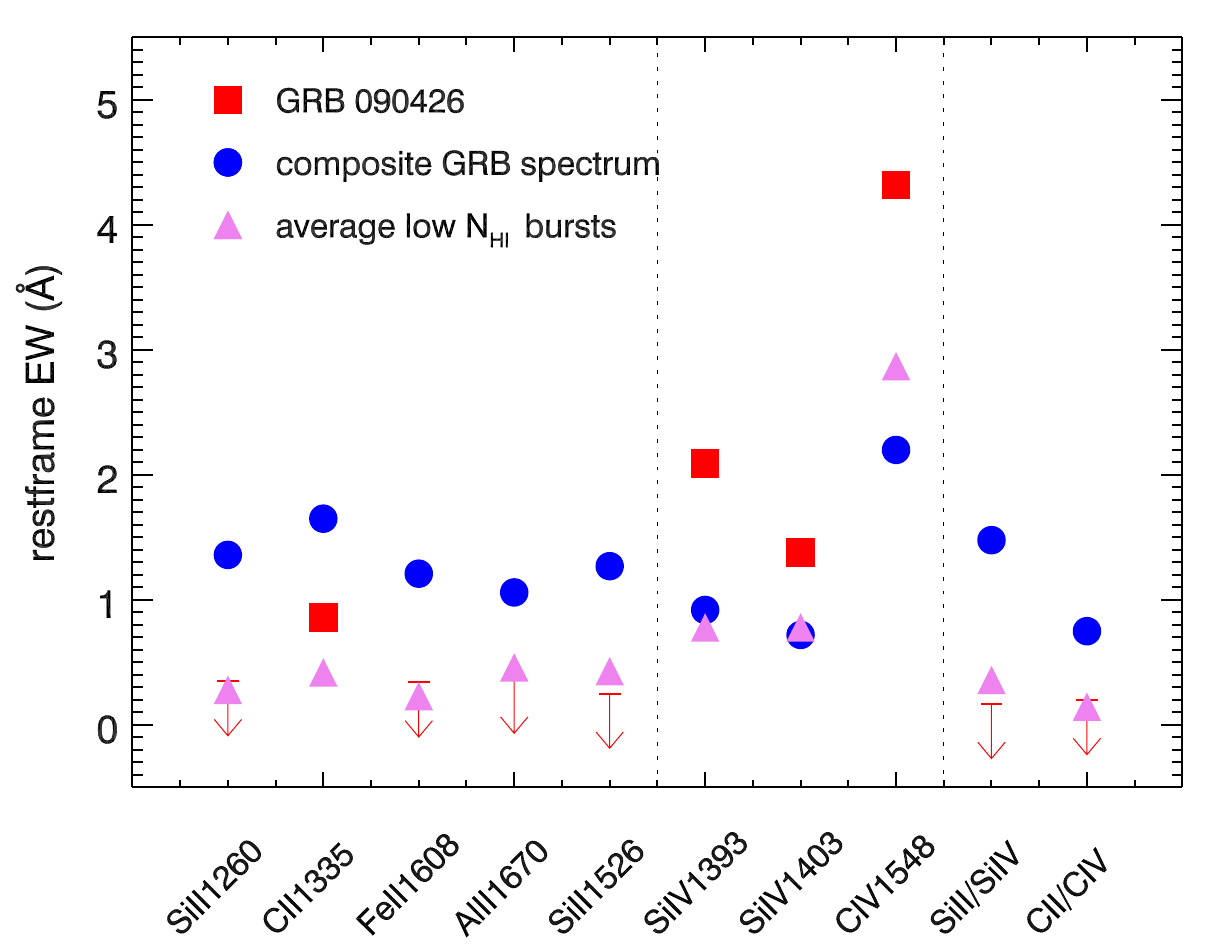}
\includegraphics[width=\columnwidth]{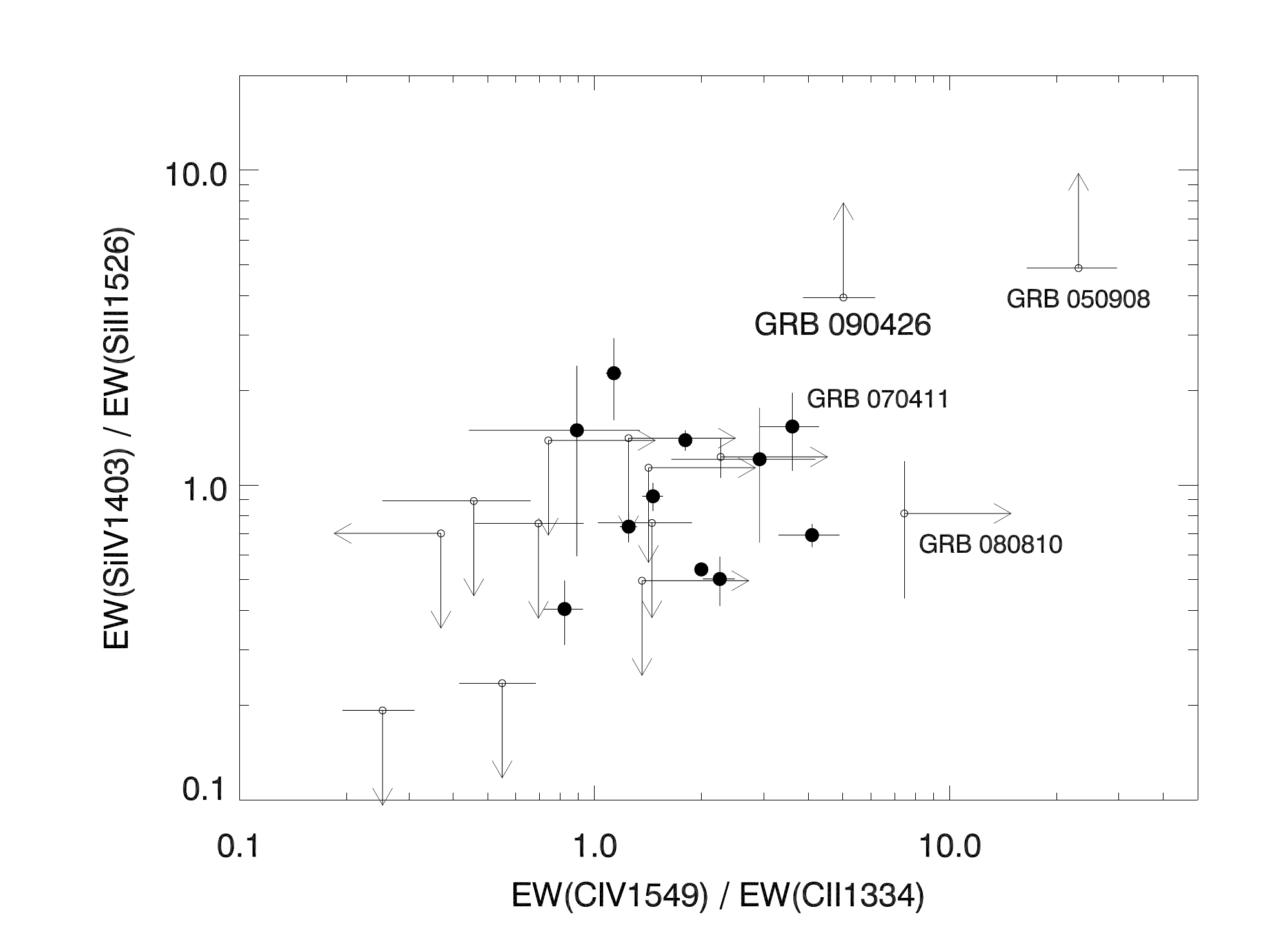}
\caption{Top: Comparison of EWs of the average optically bright long-duration {\it Swift} burst, a subsample of bursts with a low neutral hydrogen column density, and the values for GRB 090426. Also plotted are the ionization ratios of Si and C. This comparison clearly shows the exceptional line strengths of the spectrum of GRB 090426, showing a highly ionised medium. Bottom: Ionization of Si vs. C for GRB 090426 and the GRB sample presented in Fynbo et al. (2009). Listed are only those bursts with  a detection in at least one of the ionization states of Si and C. Bursts with log N$_\mathrm{HI}$/cm$^{-2}$ $<$ 19 are labeled. GRB 090426 has a high ionization for both elements, similar to other bursts with low log N$_\mathrm{HI}$.}
   \label{090426:comp}
 \end{center}
\end{figure}

If we take only a composite of sub-DLA bursts (3 spectra), we see the following: a fit of the 
Ly$\alpha$ absorption line gives log $N_\mathrm{HI}$/cm$^{-2}$=19.3$\pm$0.1 (dominated by the spectrum 
of GRB 070411). All the usual strong low ionization lines are also present in this composite such as 
Si II $\lambda$ 1260, 1304 and 1526  \AA{}, Si II* $\lambda$ 1264 \AA{},  C II $\lambda$ 1334 \AA{}, 
and Al III $\lambda$ 1670 \AA{}. Si IV $\lambda$ 1402 is not very strong, but falls in a noisy area of 
the composite, with EW$_\mathrm{rest}$= 0.25$\pm$0.20 \AA{}. In contrast, CIV 1549 is also quite strong 
here, with EW$_\mathrm{rest}$ of 3.2$\pm$0.1 \AA{}. The values in Fig. \ref{090426:comp} are 
not derived from the composite spectrum since this is dominated by the GRB 070411 spectrum; instead 
instead we took the average of the EWs of all sub-DLAs.

Comparing the ionisation of C and Si to those determined from the sample of Fynbo et al. 2009 (see Fig. \ref{090426:comp}), GRB 090426 occupies a rather different region than most burst in the sample. However, it seems that other GRBs with low N$_\mathrm{HI}$ also tend to show a rather high ionisation state of the observed medium. GRB 060607A is the most extreme case, although it is not included in the plot since only CIV has been listed in Fynbo et al. (2009). From Fig. 3 in Fox et al. (2008), who analysed the UVES spectrum of this burst, we see that this burst, together with the lowest reported log N$_\mathrm{HI}$/cm$^{-2}$ of 16.8 (Fox et al. 2008, whereas Fynbo et al. 2009 report a value of $\sim$ 17.2) for a GRB, has no C\,II or Si\,II while the high ionisation species are rather strong. GRB 050809 has relatively strong low ionisation lines, but the value in Fig. \ref{090426:comp} is influenced by the nondetection of Si\,IV $\lambda$ 1403 while the other line of the doublet at $\lambda$ 1393 is detected. This seems to confirm that there is a correlation between the hydrogen column density and the ionisation of the medium, not surprising since a large hydrogen column density would screen the UV radiation from the GRB. However, it might also imply that some bursts occur in a rather isolated region of their host galaxy, which would allow us to truly probe ionised gas from the surroundings of the burst itself.

\section{Photoionization modeling}

Photoionization models allow to us determine the distance of the observed
material from the GRB by considering the ionization state, column
densities, and their variations with time (e.g. Lazzati et
al. 2001). Here, we use the photoionisation code by Perna \& Lazzati
(2002, see also Lazzati \& Perna 2003) specifically written for
application to GRB observations. The code assumes that the GRB is
initially surrounded by a cold ($T_0=100$~K) absorbing material with
solar metallicity, either in the form of a spherical shell {\bf with a thickness of 0.1 the shell radius} or a
spherical cloud. The GRB spectrum and its temporal evolution are the input
in the code, which computes the evolution of the absorbing material's ionization state as a function of both time and the distance of the absorber from the burst. At each time, the resulting
wavelength-dependent opacity integrated over the line-of-sight is
outputted and compared to the observations.

\begin{figure}
 \begin{center}
  \includegraphics[width=\columnwidth,angle=-0]{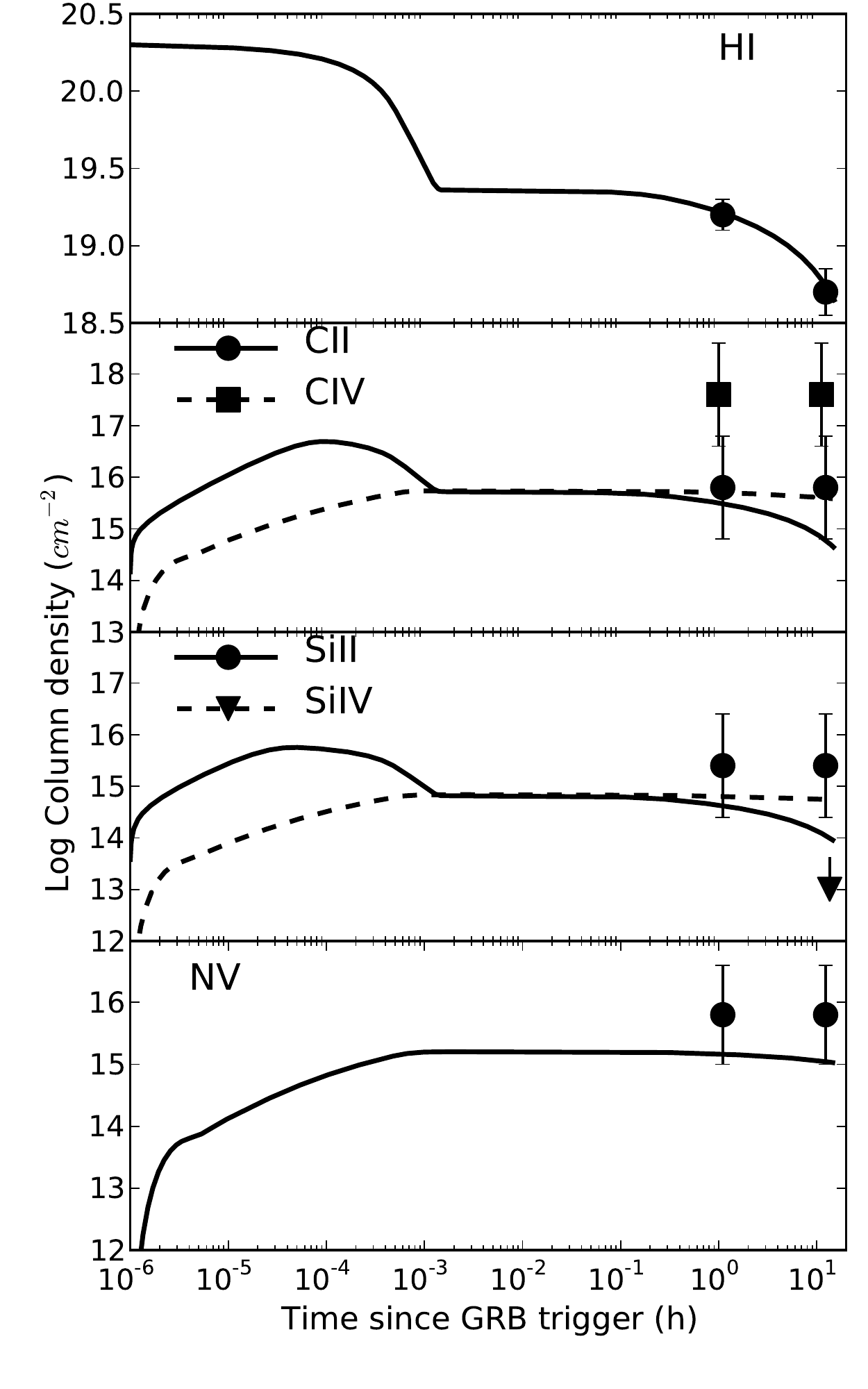}
  \caption{Temporal evolution of the absorbing columns of, from top to
    bottom, H, C, Si, and N ions. The best fit evolution of the
    absorbing columns as derived from our photoionisation model is
    overlaid.}
   \label{fig:ew}
 \end{center}
\end{figure}

The total absorption at early times is a measure of the total absorbing column, while the speed of
the evolution is a measure of the distance of the absorber. This
analysis is similar to what has been done for the combined soft-X-ray
photoionization opacity in GRB~050904 (Campana et al. 2007). GRB~090426 is better suited for this kind of analysis since
we observe variation in the HI absorption. Since hydrogen is the
most abundant element in the ISM, constraining the geometry of the
absorption region with the evolution of the Ly$\alpha$ absorption
gives us a direct result that does not depend on metallicity
considerations. The four panels of Figure~\ref{fig:ew} show the
measured column densities of the H, C, Si, and N ions with the best
absorption model overlaid. We obtain the best fit model for a uniform cloud surrounding the GRB. The
cloud properties are fit using only the HI absorption, with the other ions and
the soft-X-ray continuum absorption only used as consistency
checks. 

The absorber that best reproduces our
Ly$\alpha$ measurements is a cloud of radius $R=79$~pc (which
corresponds to the size of a large molecular cloud) and an initial HI
column density $N_{\rm{HI}}=2\times10^{20}$~cm$^{-2}$. Such values
correspond to a total mass of the molecular cloud of $\sim40,000
\msole$ and an average density $n\sim1$~cm$^{-3}$. The errors on the
values from the modeling are on the order of
$20\%$. {\bf The model does not rule out a solution in the form of a shell; however, since the thickness of the shell is 0.1R, the density would be about 10 times larger as a result. A wind blown medium would be fully ionised (see Robinson et al. 2010) at very early times and is therefore ruled out here.}

\begin{figure}
 \begin{center}
  \includegraphics[width=\columnwidth,angle=-0]{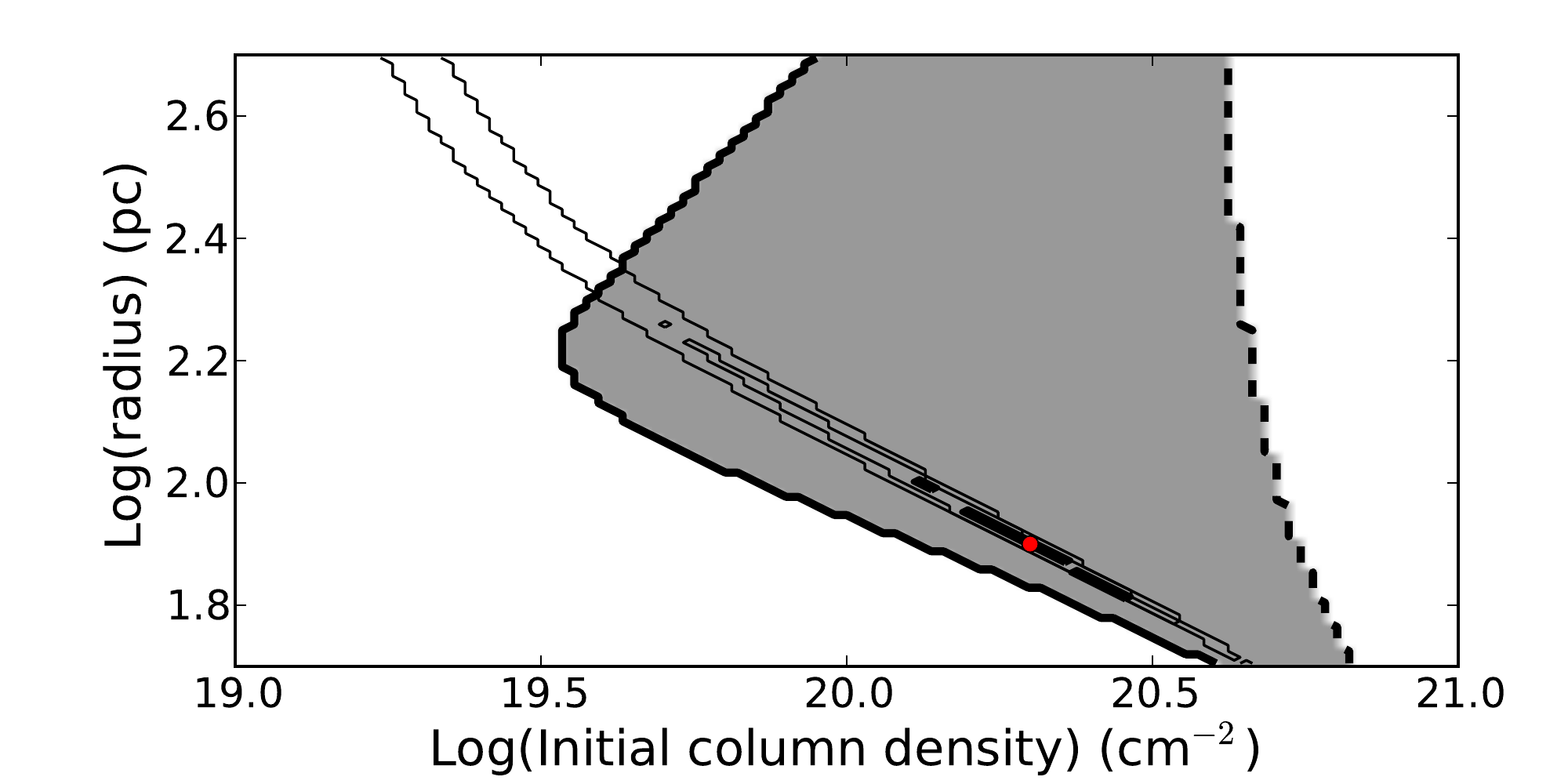}
  \caption{Constraints in the $R-N_{\rm{HI}}$ plane from the
    photoionization modeling. The thin solid lines show the 1-, 2-,
    and 3-$\sigma$ contour levels from the Ly$\alpha$ fitting
    only. The gray area shows the region allowed by the combined
    constraint on the Si\,IV evolution (right of the solid thick line, allowing only
    high column density regions) and the continuum soft X-ray
    absorption (left of the dashed line, allowing only low column
    density regions).}
   \label{fig:clev}
 \end{center}
\end{figure}

Figure~\ref{fig:clev} shows the 1-, 2-, and 3- $\sigma$ contour
levels of the Ly$\alpha$ fitting in the $R-N_{\rm{HI}}$ plane
(thin solid lines). Overlaid is the area that is
consistent with the combined constraint of Si\,IV column evolution and
soft X-ray continuum absorption. The Si\,IV constraint allows for
regions of high column density (right of the thick solid line) while
the soft X-ray absorption allows only for low column density regions
(left of the dashed line). The shaded area is in full
agreement with the Ly$\alpha$ result. A much looser
 constraint comes from the NV fitting, while the C\,IV column density cannot be fit by our model. The C\,IV excess might be an indication of a carbon-enriched medium, possibly from earlier SN explosions, but we
caution the reader due to the large errors affecting the C\,IV line (see Sec. \ref{spectrum}).

The agreement of the photoionization model predictions with the
evolution of the column densities ultimately confirms that all the
material we observe is close to the GRB site. Such material has never
been probed before in GRB afterglow spectra.

\section{The host galaxy complex}

\begin{figure}
\begin{center}
\includegraphics[width=\columnwidth]{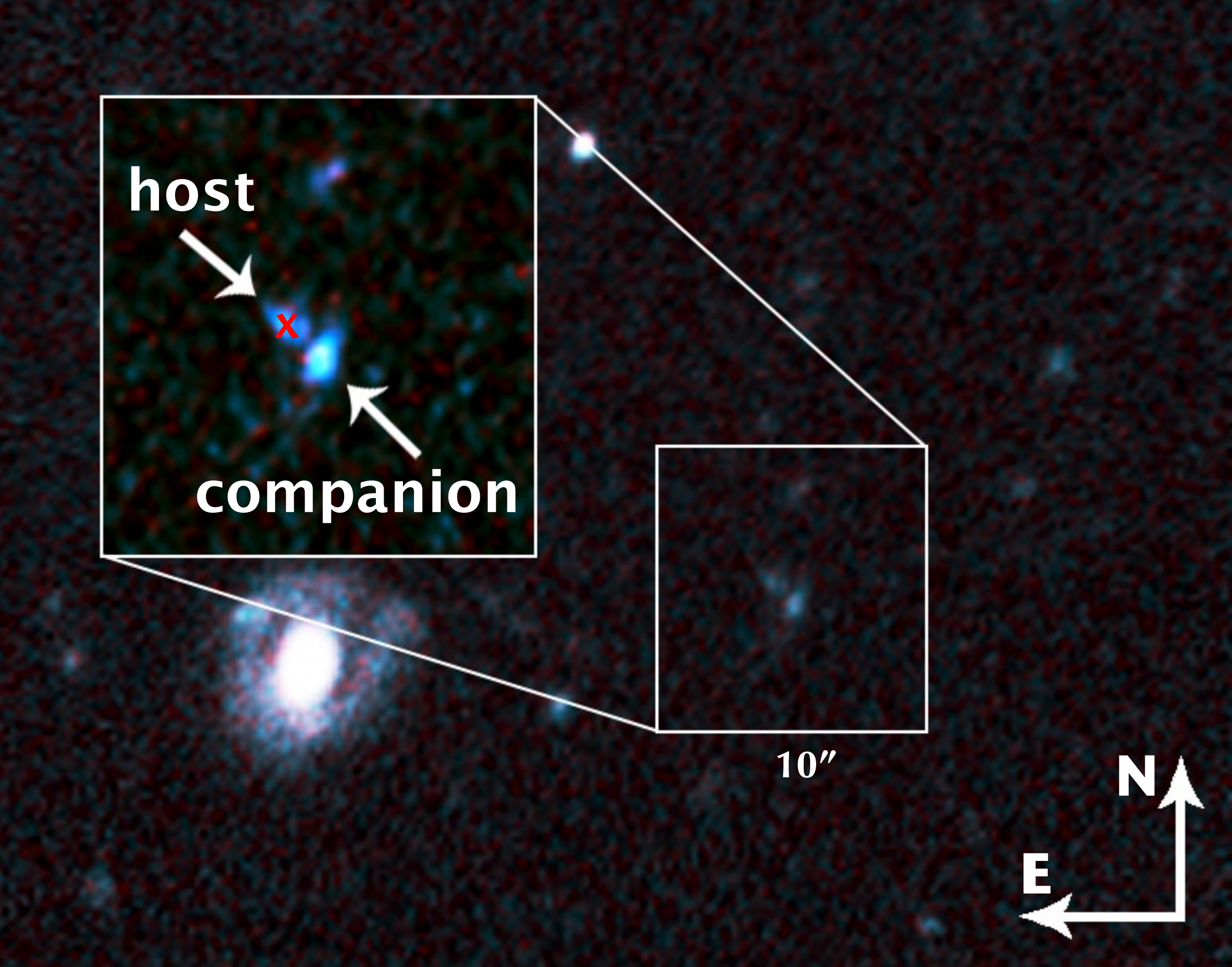}

\caption{Color image of the host galaxy field and magnified complex of the host galaxy 
(H) and the companion (G1). The position of the afterglow is marked with a red cross in the inset.}
   \label{090426:host}
 \end{center}
\end{figure}

In addition to the afterglow spectroscopy, we obtained observations of the host galaxy and its neighbour.
Using the acquisition images of the spectroscopic observations, we determined the magnitudes of 
the two galaxies G1 and H (the host) to be $V_\mathrm{G1}=24.13\pm0.05$ and $V_\mathrm{H}=24.93\pm0.08$, 
respectively, corresponding to absolute magnitudes of $M_\mathrm{AB}$(1500\,\AA{})= --21.05 and 
--20.43. At the host redshift, this implies relative luminosities of 1.5 and 0.8 $L_*$ (Reddy \& Steidel 2009). 
Fig. \ref{090426:host} shows a color image of the host galaxy complex (obtained from our 
acquisition images) and indicates the position of the afterglow. 

In the FORS2 observations, both galaxies show clear Ly$\alpha$ emission at redshifts of $z=2.605$ (H) and $2.611$ (G1), indicating that the host has about the same redshift as the GRB while the companion is at a slightly higher redshift. 
At that redshift, the physical distance between their cores 
is only 9\,kpc in projection, which implies that they are likely interacting and possibly in the 
process of merging.  The Ly$\alpha$ emission of the host galaxy gives a star formation rate of 
3.1 $M_\odot$ yr$^{-1}$, on the lower end of what is typically observed in (long duration) 
GRB hosts (Savaglio et al. 2009). 

In the 2-dimensional X-shooter spectrum we detect several faint emission lines in the near-IR, including 
[O\,III] $\lambda$ 5007, 4959 and H$\beta$. Both [O\,III] lines show two components separated by about 400 km s$^{-1}$ 
and offset in the spatial direction, which we attribute to the host (H) and G1. The second component of H$\beta$ is coincident with an atmospheric emission line. 

\begin{figure}
 \begin{center}
\includegraphics[width=\columnwidth,angle=-0]{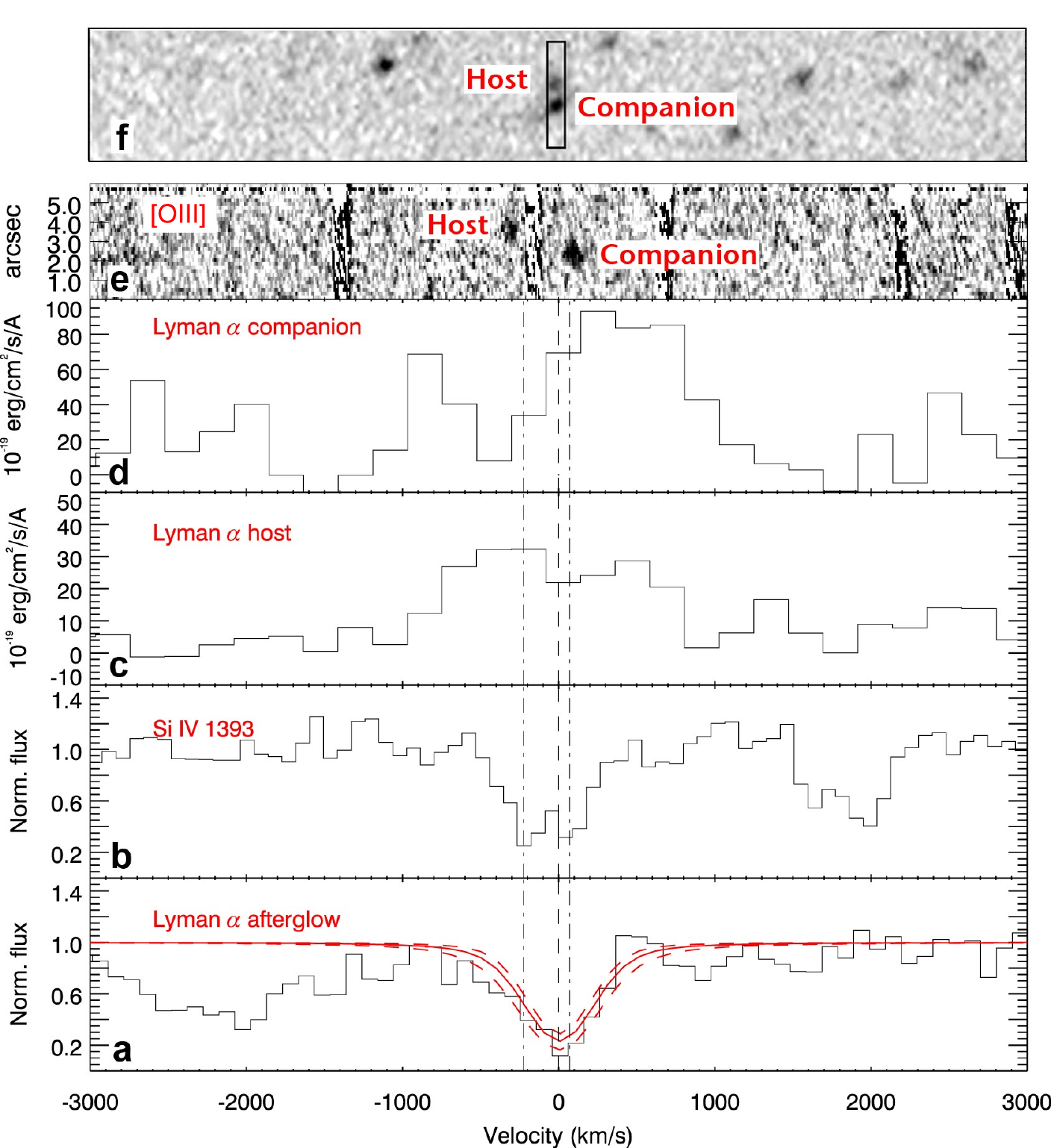}

\caption{Kinematics of the different absorption and emission lines from the GRB afterglow spectrum 
and the host galaxy complex. Panel (a): Ly$\alpha$ in absorption from the FORS2 spectra (12 hr after 
the burst) together with the fit and the 1$\sigma$ errors of the Ly$\alpha$ profile. Panel (b): 
Si\,IV in absorption with two velocity components indicated by the dashed-dotted lines. Panels (c) 
\& (d): Ly$\alpha$ in emission from the host and its companion (FORS2). Panel (e): 2D spectrum from 
X-shooter with the [O III] emission lines detected, tracing the ISM in the host and its companion. 
Panel (f): an image of the host and its companion with the slit position indicated.}
   \label{090426:kinematics}
 \end{center}
\end{figure}

Comparing the afterglow spectra and the emission lines from the hot gas of the two galaxies, the kinematics turn out to be rather complex 
(see Fig. \ref{090426:kinematics}). The Si\,IV transitions at $\lambda$ 1393 and 1403 \AA{} each 
show two equally strong velocity components with a separation of about 300 km\,s$^{-1}$. For N\,V 
and C\,IV, the two components cannot be separated since the doublets are closer together (C\,IV 
is also affected by a sky line). The red component of the Si\,IV lines is at about the same 
velocity as the center of the Ly$\alpha$ absorption. The almost equally strong blue Si IV 
component is coincident with the blue wing of Ly$\alpha$, which could be fitted with a small additional 
component at that velocity but with only 1/100 to 1/1000 the column density of the main component. 
The Ly$\alpha$ emission lines of the host and G1 are slightly redshifted compared to their nebular 
emission lines, which could imply an outflow of hydrogen gas as commonly observed in, for example, Lyman break 
galaxies (Steidel et al. 2010). Furthermore, emission and absorption line velocities from the host 
do not coincide, which suggests a complex velocity field due to the physical motion of the material seen 
in absorption. This is to be expected in the merging process of two galaxies. It also supports the 
suggestion that the material we see in absorption is actually close to the GRB and not only coming 
from the ISM of the host galaxy.

\section{Discussion and conclusions}

The spectrum of GRB 090426 is the first compelling case of observed photoionised material 
from the star-forming region of the progenitor itself. Its low hydrogen column density is varying, 
the absorption lines show a highly ionised medium of high column density, and the large discrepancy 
between X--ray and optical absorption gives another indication for photoionised material being present. 
GRB spectra have shown a large range of hydrogen column densities, usually explained by different 
sightlines of the GRB through its host galaxy (Fynbo et al. 2009; Jakobsson et al. 2006), with some 
column densities even lower than the one measured for GRB 090426. However, if GRB 090426 exploded in, for example, the halo of the galaxy or a very low density ISM like a globular cluster, we would also 
expect low column densities of the metal absorption lines. A location in the halo had been suggested 
for GRB 070125 (Cenko et al. 2008) and GRB 071003 (Perley et al. 2008), both of which show a bright but 
featureless continuum with extremely weak absorption lines. GRBs with log $N_\mathrm{HI}$/cm$^{-2}$ $<$ 20.0 tend to show a 
higher fraction of ionised versus neutral material, possibly explained by a lack of UV shielding 
by the lower density of hydrogen; however, some of them still contain a relatively large column density 
of neutral material. In addition, the afterglow showed that the GRB must have exploded in a non-low-density medium (Levesque et al. 2010). The strange properties of the spectrum of GRB 090426 
can therefore not simply be explained by a sightline effect. 

Considering the fact that this GRB exploded in what seems to be a galaxy merger, another scenario 
might explain the strange spectrum we observe. The lack of absorption from the normal ISM of the 
host suggests that the progenitor did not reside in a star-forming region inside its host galaxy. 
A possible scenario is therefore that the progenitor was located in an isolated star-forming region 
outside the host itself, e.g. in something like a tidal tail created by the merging of the two 
galaxies. Thus, the sightline to the GRB would only intersect the material of the star-forming region, 
ionised by the GRB flux and eventually piled up by the wind from massive Wolf-Rayet stars, and not 
the dense ISM of the galaxy. We note, however, that the afterglow position is, in projection, close 
to the center of the galaxy.

Another strange feature about this burst was its relatively short duration paired with a soft 
high energy emission spectrum. Naturally, the question arises if the unusual high-energy properties 
and those of the afterglow spectrum could be connected. From Fig. \ref{090426:HR} we see that all 
other low column density GRBs occupy the normal parameter space for long bursts concerning HR and 
duration, so there seems to be no obvious correlation between HR or duration and low column density. 
Since the material observed in the afterglow spectrum is relatively far away from the burst compared 
to the prompt emission, which is expected to be produced by internal shocks within the jet of 
the GRB, a connection is not to be expected. 

Concerning the progenitor of the burst, Antonelli et al. (2009) and Levesque et al. (2010) had 
already concluded that, despite its short duration, the burst was likely due to the collapse of a 
massive star. Arguments include the star-forming host galaxy, the high afterglow luminosity, consistency 
with the Amati relation, and the strength of the absorption lines. In past years, the classification 
of bursts purely on its duration or another single observational fact has also been questioned.  
Evidence for a collapsar progenitor has been observed even in the absence of a supernova signature 
(GRB 060505 and GRB 060614, see e.g. Fynbo et al. 2006b), which we cannot verify for GRB 090426 
since the SN would be too faint to detect with current instruments, and a range 
of high redshift bursts have been detected with short intrinsic duration (see e.g. L\"u et al. 2010). Suggestions for how 
to produce a collapsar with such a short intrinsic duration include viewing the GRB from off-axis, such 
that we see only a small part of the jet (Lazzati et al. 2009), or the possibility that the central engine had been 
turned off before the jet reached the surface of the str, with the destruction wave catching up with 
the jet and leading to a shorter prompt duration (Mizuta \& Aloy 2009). %done to here

No other GRB observed so far has shown such extreme properties of the absorption lines, despite a 
number of spectra with relatively low column densities. However, none of them had time resolved 
spectral observations that would allow a line variability study (with the exception of GRB 080310, a potentially similar case to GRB 090426 concerning the properties of the afterglow spectrum; see A. De Cia et al. in prep.). %In addition, a low hydrogen column density alone does not necessarily imply a similar placement of the GRB compared to its host 
%galaxy such as for GRB 090426 and in many cases this is likely to be a pure sightline effect. 
A larger number of rapid response observations and time series of spectra using a high sensitivity 
and medium- to high-resolution spectrographs (in order to fit reliable column densities) might 
allow us to study a few more of those rare events where we directly see the interaction of the 
GRB with its environment. Extremely rapid response, on the timescale of a few minutes, might even 
allow us to observe the material ejected by the progenitor itself.

\section*{Acknowledgments}

Observations are based on ESO proposals 083.D-0606(C), 083.D-0606, and commissioning data of the X-shooter 
instrument. We thank the Paranal staff, especially G. Carraro, P. Lynam, E. Mason, J. Smoker, F. J. 
Selman and S. Stefl for performing our observations. Also based on observations with the Nordic Optical Telescope under program 39-023 (PI Jakobsson), operated on the island of La Palma jointly by Denmark, Finland, Iceland, Norway, and Sweden in the Spanish Observatorio del Roque de los Muchachos of the Instituto de Astrof\'{\i}sica de Canarias. We thank the observers Zita Banhidi and S\o{}ren Frandsen for performing the NOT observations. CT and EML thank Josh Bloom for his guidance in obtaining the Keck afterglow spectrum. CT thanks Alex Kann for the discussion on the brightness of short vs. 
long afterglows. CT \& SC are partially supported by ASI grants. The Dark Cosmology Centre is funded by 
the Danish National Research Foundation. LC is supported by the DFG cluster of excellence 
``Origin and Structure of the Universe". EML's participation was funded in part by a Ford 
Foundation Predoctoral Fellowship. MAA is partly founded through grants AYA2007-67626-C03-01 
and Prometeo-2009-103.

\end{document}